%% file: mulch.tex
\documentclass[sigchi-a]{acmart}
\usepackage{booktabs} 
\usepackage{ccicons}  

\setcopyright{none}

\acmDOI{10.475/123_4}

\acmISBN{123-4567-24-567/08/06}

\acmConference[CHI 2019]{ACM CHI Conference on Human Factors in Computing Systems}{May 2019}{Glasgow, Scotland, UK}
\acmYear{2019}
\copyrightyear{2019}

\acmPrice{15.00}

\begin{document}
\title{A Mulching Proposal}
\subtitle{Analysing and Improving an Algorithmic System for Turning the Elderly into High-Nutrient Slurry}

\author{Os Keyes}
\orcid{0000-0001-5196-609X}
\affiliation{%
  \department{Department of Human Centered Design \& Engineering}
  \institution{University of Washington}
  \streetaddress{428 Sieg Hall, Box 352315}
  \city{Seattle}
  \state{WA}
  \postcode{98195-2315}
  \country{USA}}
\affiliation{Logan-Nolan Industries}
\email{okeyes@uw.edu}

\author{Jevan Hutson}
\orcid{0000-0003-3312-1733}
\affiliation{%
  \department{School of Law}
  \institution{University of Washington}
  \city{Seattle}
  \state{WA}
  \postcode{98195-2315}
  \country{USA}}
\affiliation{Logan-Nolan Industries}
  \email{jevanh@uw.edu}

\author{Meredith Durbin}
\orcid{0000-0001-7531-9815}
\affiliation{%
  \department{Department of Astronomy}
  \institution{University of Washington}
  \city{Seattle}
  \state{WA}
  \postcode{98195-2315}
  \country{USA}}
\affiliation{Logan-Nolan Industries}
  \email{mdurbin@uw.edu}

\renewcommand{\shortauthors}{Keyes \textit{et al.}}

\ccsdesc[500]{Human-centered computing~Empirical studies in HCI}
\ccsdesc[500]{Human-centered computing~Social engineering (social sciences)}
\ccsdesc[500]{Computing methodologies~Object recognition}
\ccsdesc[500]{Computing methodologies~Machine learning algorithms}
\ccsdesc[500]{Social and professional topics~Age}
\ccsdesc[300]{Applied computing~Consumer health}
\ccsdesc[500]{Applied computing~Metabolomics / metabonomics}
\ccsdesc[500]{Applied computing~Agriculture}

\begin{abstract}
The ethical implications of algorithmic systems have been much discussed in both HCI and the broader community of those interested in technology design, development and policy. In this paper, we explore the application of one prominent ethical framework\textemdash Fairness, Accountability, and Transparency\textemdash to a proposed algorithm that resolves various societal issues around food security and population ageing. Using various standardised forms of algorithmic audit and evaluation, we drastically increase the algorithm's adherence to the FAT framework, resulting in a more ethical and beneficent system. We discuss how this might serve as a guide to other researchers or practitioners looking to ensure better ethical outcomes from algorithmic systems in their line of work.
\end{abstract}

\keywords{algorithmic critique; algorithmic analysis; computer vision; dystopia; fairness; accountability; transparency; ethics}

\maketitle

\input{sections/1_introduction}

\input{sections/2_fieldsite}

\input{sections/3_results}

\input{sections/4_discussion}

\input{sections/5_conclusion}

\input{sections/6_acknowledgements}

\bibliography{mulch_refs}
\bibliographystyle{ACM-Reference-Format}

\end{document}

%% file: sections/1_introduction.tex
\section{Algorithmic Critique and HCI}

As algorithmic systems have become more widely deployed in critical social and economic domains, a range of concerns have arisen about how such systems might create unjust and harmful outcomes. These concerns have\textemdash in domains from child welfare policy to cancer treatment\textemdash often been validated~\cite{Noble2018,Eubanks2018,Wakabayashi2018,Ross2018}. 

In response, researchers have developed a range of standards by which to evaluate and critique algorithmic systems. These include not only academic proposals, such as the "Principles of Accountable Algorithms",~\cite{Diakopoulos2016b,oneill} but also proposals from major technology companies ~\cite{GooglePrin,MicroPrin,UnityPrin}. One shared set of values these standards offer can be summarised by the principles of the "Fairness, Accountability and Transparency" (FAT) ~\cite{FATML,FATCon,Lepri2017, Diakopoulos2016} framework. This says that algorithmic systems should, to be ethical, be:

\begin{enumerate}
  \item \textit{Fair}: lacking biases which create unfair and discriminatory outcomes;
  \item \textit{Accountable}: answerable to the people subject to them;
  \item \textit{Transparent}: open about how, and why, particular decisions were made.
\end{enumerate}

By assuring these conditions are met, we can rest easy, threatened no more by the possibility of an algorithm incidentally or intentionally producing harmful outcomes. 

HCI's role in this work has been to dive into unpacking these principles' meaning and developing mechanisms by which they might be met. We have developed new transparency mechanisms~\cite{Rader2018}, explored user perceptions of fairness~\cite{Lee2017}, and directly "audited" algorithms to explore whether they are biased~\cite{Robertson2018,Chen2015,Chen2018}.

But these studies are often external, undertaken by researchers who are not directly engaging with the algorithm's developers. As a consequence, researchers have a limited ability to directly correct any problems they find and test their proposed solutions. For similarly practical reasons, studies often only tackle one part of the FAT framework, focusing exclusively on (for example) transparency or fairness, rather than taking a systemic view of these principles.

In this paper, we document an algorithmic analysis which was undertaken in \textit{collaboration} with the system's developers, seeking to ensure that design followed the principles laid out in the FAT framework. Due to this collaboration, we believe that our work makes for an interesting case study in HCI engagement with industry. HCI researchers concerned about algorithmic ethics might use our experience as a basis for more efficacious and holistic engagement with the systems they are studying.

%% file: sections/2_fieldsite.tex
\section{Field Site}
\begin{marginfigure}
  \includegraphics[width=\marginparwidth]{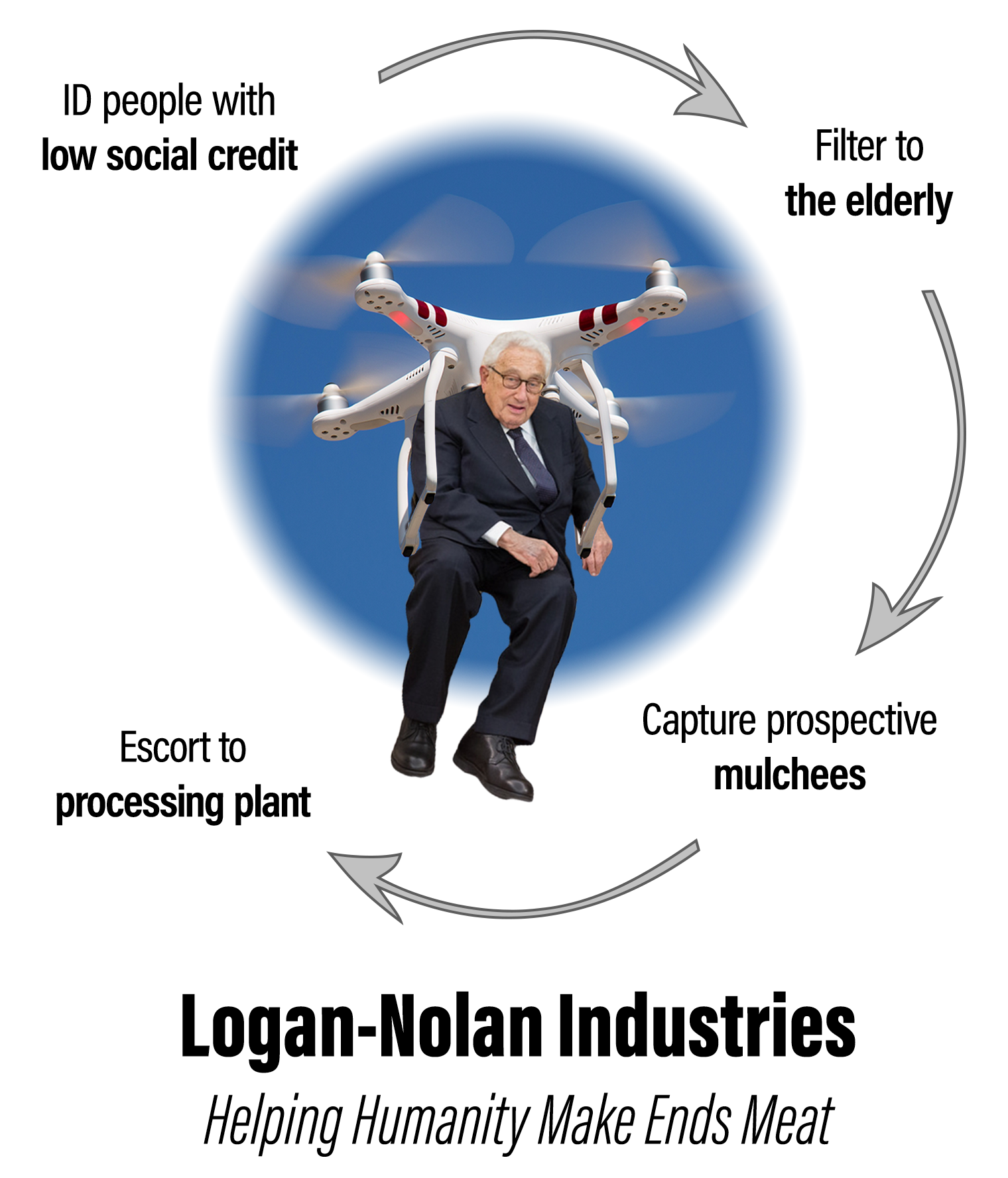}
  \caption{A publicity image for the project, produced by Logan-Nolan Industries}~\label{fig:kiss}
\end{marginfigure}

Logan-Nolan Industries (LNI) is a large multinational with a range of interests, from software development to mining. Recognising a gap in the market created by multiple ongoing crises\textemdash namely, the ongoing population aging in Western society, and the likely reduction in arable farmland due to climate change\textemdash they developed a program in which elderly people are rendered down into a fine nutrient slurry, directly addressing both issues. Elderly volunteers (or "mulchees") are provided with generous payments for their families before being rendered down and recombined with other chemicals into a range of substitutes for common foodstuffs, including hash browns (Grandmash\texttrademark), bananas (Nanas\texttrademark) and butter (Fauxghee\texttrademark).

Promisingly remunerative early trials led LNI to seek to expand\textemdash but there are only so many possible volunteers. It seems that, despite the \textit{clear} benefit the program offers humanity, many are reticent to be rendered. In an effort to move past this, LNI has developed a new, automated approach. An algorithm, provided with vast amounts of social media and communications data, identifies subjects with low levels of social connectivity, partially using a HCI-developed algorithm for approximating "social credit"~\cite{Singh2017}. Photographs of subjects who fall below a certain connectivity threshold are then run through a computer vision system which identifies those who appear to be above the age of 60. Once a person is classified as both old and unloved, their information is dispatched to a network of patrolling unmanned aerial vehicles (UAVs) who\textemdash upon spotting a person who is present in the database\textemdash obtain them and bring them to the nearest LNI processing centre.

LNI formed informal focus groups and presented them with this proposal. The company was surprised to find that possible participants responded quite negatively to the idea. LNI's expert geriatric gustatologists thus reached out to us, seeking our expertise in order to resolve the anxiety of both their consumers and those to be consumed. We were afforded full access to the development team and process, including contact with senior managers, in order to ensure our feedback could be directly implemented. The result is an interesting (and certainly unique!) case study in algorithmic systems design.

%% file: sections/3_results.tex
\section{Findings}

\subsection{Fairness}

Fairness is a complex problem\textemdash there are, after all, many definitions of "fair"~\cite{2018arXiv181107867M,corbett2018measure,foulds2018intersectional}\textemdash and we chose, as other algorithmic researchers have ~\cite{mehrotra2017auditing,Buolamwini2018}, to look specifically for \textit{demographic} fairness. Did the system (either in determining social connectivity or age) evenly distribute false positives and negatives across racial and gender groups?

We assembled a dataset of 900 images across these demographic axes, sourced from LNI employees and family members who consented (through their employment contracts) to allow us to assess their social credit score and use their photographs. Having quickly skimmed Keyes's \textit{The Misgendering Machines}~\cite{keyes2018misgendering}, we saw a need to include transgender (trans) people in our dataset, and expanded our model of gender in order to do so. The resulting data was tested against the LNI algorithm; our results can be seen in table 1.

\begin{table}[H]
  \caption{Percentage of individuals tagged as worthy of mulching, by demographic.}
  \label{tab:table1}
  \begin{tabular}{l r r r r r}
    & & \multicolumn{2}{c}{\small{\textbf{Mulching Probability}}} \\
    \cmidrule(r){3-4}
    {\small\textit{Race}}
    & {\small \textit{Cis Man}}
      & {\small \textit{Cis Woman}}
    & {\small \textit{Trans Man}}
    & {\small \textit{Trans Woman}}
    & {\small \textit{Non-Binary Person}} \\
    \midrule
    White & 44.6\% & 33.3\% & 2.2\% & 3.2\% & 1.1\% \\
    Asian-American & 22.2\% & 16.3\% & 2.8\% & 1.2\% & 1.8\% \\
    African-American & 26.9\% & 11.2\% & 2.3\% & 1.9\% & 3.4\% \\
    Latino & 16.9\% & 18.7\% & 3.3\% & 1.2\% & 1.7\% \\
    Native American & 14.4\% & 12.4\% & 1.0\% & 0.8\% & 1.5\% \\
    Hawaiian \& Pacific Islander & 11.6\% & 7.8\% & 2.4\% & 1.1\% & 0.7\%\\
     \bottomrule
  \end{tabular}
\end{table}

As shown, the algorithm disproportionately tagged white cisgender men as worthy of mulching, biasing against other populations. While we cannot adequately explain what wider reason the algorithm might have for determining that white cisgender men are, on average, lacking in societal worth, this is clearly unacceptable. For an algorithm to be fair, it must lack gender or racial bias.

We provided our results and concerns to LNI's engineers, who were eager (unsurprisingly, given the demographics of the average engineering department) to address this issue. They responded by collecting the photographs and social credit data of 3,000 more potential mulchees, particularly women, trans people and/or people of colour. These images and data traces were integrated into the model, which (as seen in Table 2) now produces far fairer results.

\begin{table}[H]
  \caption{Post-audit mulching probabilities.}
  \label{tab:table2}
  \begin{tabular}{l r r r r r}
    & & \multicolumn{2}{c}{\small{\textbf{Mulching Probability}}} \\
    \cmidrule(r){3-4}
    {\small\textit{Race}}
    & {\small \textit{Cis Man}}
      & {\small \textit{Cis Woman}}
    & {\small \textit{Trans Man}}
    & {\small \textit{Trans Woman}}
    & {\small \textit{Non-Binary Person}} \\
    \midrule
    White & 44.6\% & 43.3\% & 44.2\% & 46.3\% & 41.2\% \\
    Asian-American & 52.2\% & 51.3\% & 55.8\% & 49.6\% & 52.3\% \\
    African-American & 46.9\% & 51.1\% & 53.2\% & 49.1\% & 53.3\% \\
    Latino & 56.9\% & 48.2\% & 47.3\% & 51.1\% & 47.4\% \\
    Native American & 54.4\% & 54.2\% & 51.5\% & 48.8\% & 51.2\% \\
    Hawaiian \& Pacific Islander & 51.6\% & 48.6\% & 44.9\% & 51.1\% & 47.0\%\\
     \bottomrule
  \end{tabular}
\end{table}
\subsection{Accountability}

\begin{sidebar}
  \textbf{User Feedback}
\vspace{5mm}
\begin{quote}
\textit{"I don't know if I'm comfortable eating Nonna"}\\ \textbf{Judith}, grand-daughter of a potential mulchee.
\end{quote}
\vspace{5mm}

\begin{quote}
\textit{"Until that little robot showed up I'd never even heard of this program. Say, how did you get in my house, anyway?"}\\ \textbf{Robert}, a potential mulchee classified as "not to be mulched".
\end{quote}
\vspace{5mm}

\begin{quote}
\textit{"Do the papers know about this kind of thing? There ought to be some investigation!"}\\ \textbf{Joan}, a potential mulchee classified as "to be mulched".
\end{quote}
\vspace{5mm}

\begin{quote}
\textit{"Ow!"}\\ \textbf{Colin}, being mulched.
\end{quote}
\end{sidebar}

Accountability is a vital component of the FAT framework, and refers to the ability of people to address any failure or inaccuracy produced by an algorithmic system~\cite{Neyland2015fk}. This is particularly important here given the consequences of an incorrect classification.

In the case of this algorithm, a failure necessitating accountability might happen at various points. The computer vision algorithm itself could fail, incorrectly classifying someone as elderly, or the analysis of social connections might be inaccurate due to a person's limited presence on social media sites.

Initial versions of LNI's algorithm lacked mechanisms for accountability\textemdash they were not responsive to possible user concerns, not even leaving as much as a voicemail, let alone providing any mechanism of redress if someone felt that they (or more likely, their friend or relative) had been rendered into slurry incorrectly.

To address accountability concerns, we undertook a round of \textit{formalised} user testing, soliciting feedback from mulchees and their relatives and friends at various stages in the mulching process. Some examples of the feedback can be seen in the sidebar. Based on the feedback, we proposed two mechanisms for accountability\textemdash one appearing prior to mulching, and the other after, and both interlinking with concerns around \textit{Transparency} (see below).

The pre-mulching accountability mechanism involves the drone itself. After approaching a pending mulchee, the drone informs them that they have been selected for participation in this program. They are then afforded a ten-second window in which to state whether they feel their selection was correct or not. In the event that they feel they were selected in error, they are put through to a human operator at LNI customer services. The operator ("or death doula") discusses the reasons behind the customer's classification, and presents them with an opportunity to discuss possible factors in age or societal utility the algorithm may have overlooked. They then either confirm or reverse the algorithm's decision. Their decisions are reported back to the algorithmic development team for consideration when adding new variables to the model or altering variable weight.

Post-mulching, the company reaches out to the friends and family of the mulchee (if such individuals exist) to inform them of the decision reached and provide the serial numbers of any food products containing their relative. Our user studies showed that people express some qualms about eating their grandparents. In the event that next-of-kin feel the decision was made wrongly, they are offered a 30-day window in which to appeal. While LNI cannot reconstitute their loved one, the company has agreed to provide an elderly person of equal or greater wholesomeness and social utility, at discounted cost.

While proposing these ideas we were highly cognizant of potential cultural bias: the 10-second and 30-day window are clearly highly variable periods of time in the context of language proficiency. For those reasons, the company has invested vast sums in accumulating additional data on not only possible mulchees but also their social contexts, ensuring that any letters and/or employees can engage in the customer's native language.

\subsection{Transparency}

Deeply intertwined with \textit{Accountability} is the principle of \textit{Transparency}: are users aware of how the algorithm reaches decisions? This is vital in allowing users and regulators to evaluate any algorithmic system.

We address transparency in several ways. During the pre-mulching accountability mechanism, the drone provides not only the decision but also a comprehensive list of variables that were considered\textemdash including but not limited to phone and SMS metadata, number of facebook friends, number of birthday and christmas cards received from relatives\textemdash along with the scores the mulchee received \textit{for} each variable. These are also (albeit by letter) provided to those who narrowly fail to meet the threshold for mulching.

LNI executives were particularly enthused at participants' comments (see the user feedback in the sidebar) that the "not mulched" notice made them far more aware of the program, raising public attention and serving as free advertising. Building upon this, and in the name of transparency, we have encouraged LNI to provide additional notice and achieve more rigorous consent from potential participants by deploying ubiquitous signage in public and private settings \cite{Burgess}. This can be seen as Figure 2.

\begin{marginfigure}
  \includegraphics[width=\marginparwidth]{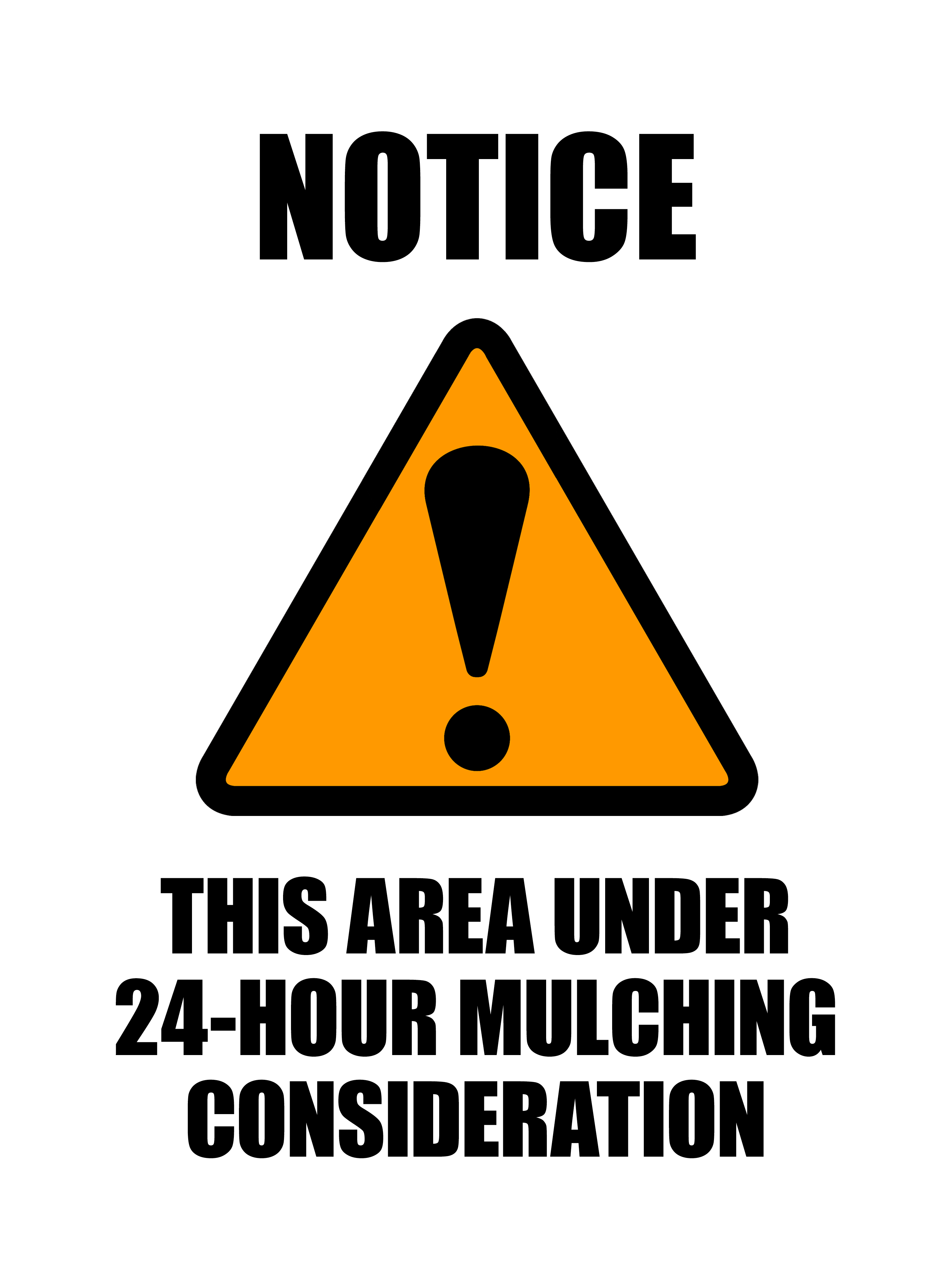}
  \caption{A warning sign of drones' presence, serving as a mechanism for consent.}
  \label{fig:sample}
\end{marginfigure}

In undertaking this work we were cognizant of the ethical constraints that commercial contexts often generate. In particular, the proprietary nature of algorithmic systems is often a roadblock for algorithmic transparency~\cite{brauneis2018algorithmic,wexler2018life}. LNI felt it was important overcome this obstacle in order to promote broader understanding of its system (as well as gain competitive advantage over other providers). Its solution has been to actively waive any right to legally pursue third-party researchers who audit or otherwise examine LNI's software.

We also suggested\textemdash in line with the feedback we gathered through user testing\textemdash that there was some advantage in being more proactive in public audit and transparency efforts. Our proposal was to build tools that would allow members of the public to easily interact with the model. The result is an open website (www.mulchme.com) where members of the public can upload their own photographs and data, play around with the model's features, and see what results. LNI saw this as not only a boost to transparency, but also an opportunity to collect further data for testing and refining their model.

%% file: sections/4_discussion.tex
\section{Discussion}

We realise that some researchers (be they potential mulchees or not) might react to this paper with some disquiet. In particular, one might argue that the paper undermines itself: that rather than showing how useful Fairness, Accountability and Transparency can be in reducing the negative possibilities of an algorithm, it shows its paucity.

In particular, such researchers could point to the fact that this work assumes algorithmic failures to be a question of implementation. Issues with the morality of this algorithm are treated as resolvable through changes to input data and deployment. Such a frame of ethics ignores whether murdering the elderly might be morally obscene in principle. Sometimes the problem is not how the sausage gets made, but that they're making people into sausage.

One could also argue that, as might be expected from this tacit acceptance of the datafication of society, our ethical frame defers to the concerns and priorities of Logan-Nolan Industries and their (neoliberal) motivations. From that view, our direct collaboration with LNI is not to be lauded, but instead condemned as legitimising their actions with the stamp of scientific approval, and accepting these inhumane infrastructures as inevitable. We are collaborators not in the sense of team-mates, but in the sense of Vichy France.

Furthermore, even were we not complicit in this legitimisation, our approach fundamentally centers the treatment of ethics as a series of heuristic checkboxes that can be resolved technically, or through design methods, rather than as representative of wider societal issues. In doing so, it simultaneously over-fits (by ignoring the role that context plays in algorithmic propriety) and under-fits (by ignoring how these systems play into existing inequities)~\cite{greene2018better,Browne2015vz}.

Rather than run interference for Logan-Nolan Industries, they might say, we should instead begin challenging them directly, speaking out against \textit{any} form of grandma-mashing machine, whether it is driven by an algorithm or a human. Perhaps we could consider developing and applying the same critical thinking skills when \textit{evaluating} this algorithm that data scientists are being encouraged to apply when \textit{developing} them:

\begin{quote}
\textit{"Would the implementation of this algorithm represent a reformist or non-reformist reform? Is the design of this algorithm affirmative or critical? Would empowering our project partner with this algorithm entrench or challenge oppression and inequality? Such efforts can help data scientists interrogate their notions of "good" to engage in non-reformist, critical, and anti-oppressive data science."}~\cite{2018arXiv181103435G}
\end{quote}

But we are simply researchers\textemdash researchers seeking a fairer, more transparent and more accountable world, for the benefit of all. If this framing is insufficient, well: that would imply FAT \textit{itself} were insufficient. And were \textit{that} the case, it would say some pretty negative things about HCI's engagement with algorithmic evaluation, and carry some profoundly depressing implications about the actual meaningful impact we are likely to make~\cite{greene2018better}.

%% file: sections/5_conclusion.tex
\section{Conclusion}

In this paper we have presented a case study in algorithmic analysis, centred around a system that we hope will take big bites out of both food insecurity and population imbalances. Secure in the knowledge that nothing data ethicists would ask of us has been missed, we are excited to see what other researchers make of our techniques as we kick back, put our feet up, and enjoy a nice big plate of Fauxghee\texttrademark.

%% file: sections/6_acknowledgements.tex
\section{Acknowledgements}
We are extremely grateful to Meg Drouhard, Margret Wander, Vincente O. Rafael, Erin Thayer, Helena Edwards, Rachel Ingram, Chaya Helstrom and Masya Billig for their ongoing support. No Kissingers were harmed in the making of this paper.

%% file: mulch.bbl

\begin{thebibliography}{32}


\ifx \showCODEN    \undefined \def \showCODEN     #1{\unskip}     \fi
\ifx \showDOI      \undefined \def \showDOI       #1{#1}\fi
\ifx \showISBNx    \undefined \def \showISBNx     #1{\unskip}     \fi
\ifx \showISBNxiii \undefined \def \showISBNxiii  #1{\unskip}     \fi
\ifx \showISSN     \undefined \def \showISSN      #1{\unskip}     \fi
\ifx \showLCCN     \undefined \def \showLCCN      #1{\unskip}     \fi
\ifx \shownote     \undefined \def \shownote      #1{#1}          \fi
\ifx \showarticletitle \undefined \def \showarticletitle #1{#1}   \fi
\ifx \showURL      \undefined \def \showURL       {\relax}        \fi
\providecommand\bibfield[2]{#2}
\providecommand\bibinfo[2]{#2}
\providecommand\natexlab[1]{#1}
\providecommand\showeprint[2][]{arXiv:#2}

\bibitem[\protect\citeauthoryear{Brauneis and Goodman}{Brauneis and
  Goodman}{2018}]%
        {brauneis2018algorithmic}
\bibfield{author}{\bibinfo{person}{Robert Brauneis} {and}
  \bibinfo{person}{Ellen~P Goodman}.} \bibinfo{year}{2018}\natexlab{}.
\newblock \showarticletitle{Algorithmic transparency for the smart city}.
\newblock \bibinfo{journal}{\emph{Yale JL \& Tech.}}  \bibinfo{volume}{20}
  (\bibinfo{year}{2018}), \bibinfo{pages}{103}.
\newblock


\bibitem[\protect\citeauthoryear{Browne}{Browne}{2015}]%
        {Browne2015vz}
\bibfield{author}{\bibinfo{person}{Simone Browne}.}
  \bibinfo{year}{2015}\natexlab{}.
\newblock \bibinfo{booktitle}{\emph{{Dark Matters}}}.
\newblock \bibinfo{publisher}{Duke University Press}.
\newblock


\bibitem[\protect\citeauthoryear{Buolamwini and Gebru}{Buolamwini and
  Gebru}{2018}]%
        {Buolamwini2018}
\bibfield{author}{\bibinfo{person}{Joy Buolamwini} {and}
  \bibinfo{person}{Timnit Gebru}.} \bibinfo{year}{2018}\natexlab{}.
\newblock \showarticletitle{{Gender Shades: Intersectional Accuracy Disparities
  in Commercial Gender Classification}}.
\newblock \bibinfo{journal}{\emph{Conference on Fairness, Accountability and
  Transparency}}  \bibinfo{volume}{81} (\bibinfo{year}{2018}),
  \bibinfo{pages}{1--15}.
\newblock


\bibitem[\protect\citeauthoryear{Burgess}{Burgess}{2018}]%
        {Burgess}
\bibfield{author}{\bibinfo{person}{Matt Burgess}.}
  \bibinfo{year}{2018}\natexlab{}.
\newblock \bibinfo{title}{Twitter}.
\newblock
\newblock
\urldef\tempurl%
\url{https://twitter.com/mattburgess1/status/1074731539030786048}
\showURL{%
Retrieved December 24, 2018 from \tempurl}


\bibitem[\protect\citeauthoryear{Chen, Ma, Hann{\'a}k, and Wilson}{Chen
  et~al\mbox{.}}{2018}]%
        {Chen2018}
\bibfield{author}{\bibinfo{person}{Le Chen}, \bibinfo{person}{Ruijun Ma},
  \bibinfo{person}{Anik{\'o} Hann{\'a}k}, {and} \bibinfo{person}{Christo
  Wilson}.} \bibinfo{year}{2018}\natexlab{}.
\newblock \showarticletitle{{Investigating the Impact of Gender on Rank in
  Resume Search Engines}}. In \bibinfo{booktitle}{\emph{the 2018 CHI
  Conference}}. \bibinfo{publisher}{ACM Press}, \bibinfo{address}{New York, New
  York, USA}, \bibinfo{pages}{1--14}.
\newblock


\bibitem[\protect\citeauthoryear{Chen, Mislove, and Wilson}{Chen
  et~al\mbox{.}}{2015}]%
        {Chen2015}
\bibfield{author}{\bibinfo{person}{Le Chen}, \bibinfo{person}{Alan Mislove},
  {and} \bibinfo{person}{Christo Wilson}.} \bibinfo{year}{2015}\natexlab{}.
\newblock \showarticletitle{{Peeking Beneath the Hood of Uber}}. In
  \bibinfo{booktitle}{\emph{the 2015 ACM Conference}}. \bibinfo{publisher}{ACM
  Press}, \bibinfo{address}{New York, New York, USA},
  \bibinfo{pages}{495--508}.
\newblock


\bibitem[\protect\citeauthoryear{Corbett-Davies and Goel}{Corbett-Davies and
  Goel}{2018}]%
        {corbett2018measure}
\bibfield{author}{\bibinfo{person}{Sam Corbett-Davies} {and}
  \bibinfo{person}{Sharad Goel}.} \bibinfo{year}{2018}\natexlab{}.
\newblock \showarticletitle{The measure and mismeasure of fairness: A critical
  review of fair machine learning}.
\newblock \bibinfo{journal}{\emph{arXiv preprint arXiv:1808.00023}}
  (\bibinfo{year}{2018}).
\newblock


\bibitem[\protect\citeauthoryear{Corporation}{Corporation}{2018}]%
        {MicroPrin}
\bibfield{author}{\bibinfo{person}{Microsoft Corporation}.}
  \bibinfo{year}{2018}\natexlab{}.
\newblock \bibinfo{title}{Our Approach to AI}.
\newblock
\newblock
\urldef\tempurl%
\url{https://www.microsoft.com/en-us/ai/our-approach-to-ai}
\showURL{%
Retrieved December 19, 2018 from \tempurl}


\bibitem[\protect\citeauthoryear{Diakopoulos}{Diakopoulos}{2016}]%
        {Diakopoulos2016}
\bibfield{author}{\bibinfo{person}{Nicholas Diakopoulos}.}
  \bibinfo{year}{2016}\natexlab{}.
\newblock \showarticletitle{{Accountability in algorithmic decision making}}.
\newblock \bibinfo{journal}{\emph{Commun. ACM}} \bibinfo{volume}{59},
  \bibinfo{number}{2} (\bibinfo{date}{Jan.} \bibinfo{year}{2016}),
  \bibinfo{pages}{56--62}.
\newblock


\bibitem[\protect\citeauthoryear{Diakopoulos, Friedler, Arenas, Barocas, Hay,
  Howe, Jagadish, Unsworth, Sahuguet, Venkatasubramanian, Wilson, Yu, and
  Zevenbergen}{Diakopoulos et~al\mbox{.}}{2018}]%
        {Diakopoulos2016b}
\bibfield{author}{\bibinfo{person}{Nicholas Diakopoulos},
  \bibinfo{person}{Sorelle Friedler}, \bibinfo{person}{Marcelo Arenas},
  \bibinfo{person}{Solon Barocas}, \bibinfo{person}{Michael Hay},
  \bibinfo{person}{Bill Howe}, \bibinfo{person}{H.V. Jagadish},
  \bibinfo{person}{Kris Unsworth}, \bibinfo{person}{Arnaud Sahuguet},
  \bibinfo{person}{Suresh Venkatasubramanian}, \bibinfo{person}{Christo
  Wilson}, \bibinfo{person}{Cong Yu}, {and} \bibinfo{person}{Bendert
  Zevenbergen}.} \bibinfo{year}{2018}\natexlab{}.
\newblock \bibinfo{title}{Principles for Accountable Algorithms and a Social
  Impact Statement for Algorithms}.
\newblock
\newblock
\urldef\tempurl%
\url{https://www.fatml.org/resources/principles-for-accountable-algorithms}
\showURL{%
Retrieved October 28, 2018 from \tempurl}


\bibitem[\protect\citeauthoryear{Eubanks}{Eubanks}{2018}]%
        {Eubanks2018}
\bibfield{author}{\bibinfo{person}{Virginia Eubanks}.}
  \bibinfo{year}{2018}\natexlab{}.
\newblock \bibinfo{booktitle}{\emph{{Automating Inequality}}}.
\newblock \bibinfo{publisher}{St. Martin{\textquoteright}s Press}.
\newblock


\bibitem[\protect\citeauthoryear{FAT}{FAT}{2018}]%
        {FATCon}
\bibfield{author}{\bibinfo{person}{ACM FAT}.} \bibinfo{year}{2018}\natexlab{}.
\newblock \bibinfo{title}{ACM Conference on Fairness, Accountability, and
  Transparency (ACM FAT*)}.
\newblock
\newblock
\urldef\tempurl%
\url{https://www.fatconference.org/}
\showURL{%
Retrieved December 19, 2018 from \tempurl}


\bibitem[\protect\citeauthoryear{Foulds and Pan}{Foulds and Pan}{2018}]%
        {foulds2018intersectional}
\bibfield{author}{\bibinfo{person}{James Foulds} {and} \bibinfo{person}{Shimei
  Pan}.} \bibinfo{year}{2018}\natexlab{}.
\newblock \showarticletitle{An Intersectional Definition of Fairness}.
\newblock \bibinfo{journal}{\emph{arXiv preprint arXiv:1807.08362}}
  (\bibinfo{year}{2018}).
\newblock


\bibitem[\protect\citeauthoryear{Green}{Green}{2018}]%
        {2018arXiv181103435G}
\bibfield{author}{\bibinfo{person}{Ben Green}.}
  \bibinfo{year}{2018}\natexlab{}.
\newblock \showarticletitle{{Data Science as Political Action - Grounding Data
  Science in a Politics of Justice.}}
\newblock \bibinfo{journal}{\emph{CoRR}}  \bibinfo{volume}{1811}
  (\bibinfo{year}{2018}), \bibinfo{pages}{arXiv:1811.03435}.
\newblock


\bibitem[\protect\citeauthoryear{Greene, Hoffmann, and Stark}{Greene
  et~al\mbox{.}}{[n. d.]}]%
        {greene2018better}
\bibfield{author}{\bibinfo{person}{Daniel Greene}, \bibinfo{person}{Anna~Lauren
  Hoffmann}, {and} \bibinfo{person}{Luke Stark}.} \bibinfo{year}{[n.
  d.]}\natexlab{}.
\newblock \showarticletitle{Better, Nicer, Clearer, Fairer: A Critical
  Assessment of the Movement for Ethical Artificial Intelligence and Machine
  Learning}.
\newblock


\bibitem[\protect\citeauthoryear{Keyes}{Keyes}{2018}]%
        {keyes2018misgendering}
\bibfield{author}{\bibinfo{person}{Os Keyes}.} \bibinfo{year}{2018}\natexlab{}.
\newblock \showarticletitle{The Misgendering Machines: Trans/HCI Implications
  of Automatic Gender Recognition}.
\newblock \bibinfo{journal}{\emph{Proceedings of the ACM on Human-Computer
  Interaction}} \bibinfo{volume}{2}, \bibinfo{number}{CSCW}
  (\bibinfo{year}{2018}), \bibinfo{pages}{88}.
\newblock


\bibitem[\protect\citeauthoryear{Lee and Baykal}{Lee and Baykal}{2017}]%
        {Lee2017}
\bibfield{author}{\bibinfo{person}{Min~Kyung Lee} {and} \bibinfo{person}{Su
  Baykal}.} \bibinfo{year}{2017}\natexlab{}.
\newblock \showarticletitle{{Algorithmic Mediation in Group Decisions}}. In
  \bibinfo{booktitle}{\emph{ACM Conference on Computer Supported Cooperative
  Work}}. \bibinfo{publisher}{ACM Press}, \bibinfo{address}{New York, New York,
  USA}, \bibinfo{pages}{1035--1048}.
\newblock


\bibitem[\protect\citeauthoryear{Lepri, Oliver, Letouz{\'e}, Pentland, and
  Vinck}{Lepri et~al\mbox{.}}{2017}]%
        {Lepri2017}
\bibfield{author}{\bibinfo{person}{Bruno Lepri}, \bibinfo{person}{Nuria
  Oliver}, \bibinfo{person}{Emmanuel Letouz{\'e}}, \bibinfo{person}{Alex
  Pentland}, {and} \bibinfo{person}{Patrick Vinck}.}
  \bibinfo{year}{2017}\natexlab{}.
\newblock \showarticletitle{{Fair, Transparent, and Accountable Algorithmic
  Decision-making Processes}}.
\newblock  (\bibinfo{date}{Aug.} \bibinfo{year}{2017}), \bibinfo{pages}{1--17}.
\newblock


\bibitem[\protect\citeauthoryear{Mehrotra, Anderson, Diaz, Sharma, Wallach, and
  Yilmaz}{Mehrotra et~al\mbox{.}}{2017}]%
        {mehrotra2017auditing}
\bibfield{author}{\bibinfo{person}{Rishabh Mehrotra}, \bibinfo{person}{Ashton
  Anderson}, \bibinfo{person}{Fernando Diaz}, \bibinfo{person}{Amit Sharma},
  \bibinfo{person}{Hanna Wallach}, {and} \bibinfo{person}{Emine Yilmaz}.}
  \bibinfo{year}{2017}\natexlab{}.
\newblock \showarticletitle{Auditing search engines for differential
  satisfaction across demographics}. In \bibinfo{booktitle}{\emph{Proceedings
  of the 26th International Conference on World Wide Web Companion}}.
  International World Wide Web Conferences Steering Committee,
  \bibinfo{pages}{626--633}.
\newblock


\bibitem[\protect\citeauthoryear{Mitchell, Potash, and Barocas}{Mitchell
  et~al\mbox{.}}{2018}]%
        {2018arXiv181107867M}
\bibfield{author}{\bibinfo{person}{Shira Mitchell}, \bibinfo{person}{Eric
  Potash}, {and} \bibinfo{person}{Solon Barocas}.}
  \bibinfo{year}{2018}\natexlab{}.
\newblock \showarticletitle{{Prediction-Based Decisions and Fairness: A
  Catalogue of Choices, Assumptions, and Definitions}}.
\newblock \bibinfo{journal}{\emph{arXiv.org}} (\bibinfo{date}{Nov.}
  \bibinfo{year}{2018}), \bibinfo{pages}{arXiv:1811.07867}.
\newblock
\showeprint[arxiv]{1811/1811.07867}


\bibitem[\protect\citeauthoryear{ML}{ML}{2018}]%
        {FATML}
\bibfield{author}{\bibinfo{person}{FAT ML}.} \bibinfo{year}{2018}\natexlab{}.
\newblock \bibinfo{title}{Fairness, Accountability, and Transparency in Machine
  Learning}.
\newblock
\newblock
\urldef\tempurl%
\url{http://www.fatml.org/}
\showURL{%
Retrieved December 24, 2018 from \tempurl}


\bibitem[\protect\citeauthoryear{Neyland}{Neyland}{2015}]%
        {Neyland2015fk}
\bibfield{author}{\bibinfo{person}{Daniel Neyland}.}
  \bibinfo{year}{2015}\natexlab{}.
\newblock \showarticletitle{{Bearing Account-able Witness to the Ethical
  Algorithmic System}}.
\newblock \bibinfo{journal}{\emph{Science, Technology, {\&} Human Values}}
  \bibinfo{volume}{41}, \bibinfo{number}{1} (\bibinfo{date}{Sept.}
  \bibinfo{year}{2015}), \bibinfo{pages}{50--76}.
\newblock


\bibitem[\protect\citeauthoryear{Noble}{Noble}{2018}]%
        {Noble2018}
\bibfield{author}{\bibinfo{person}{Safiya~Umoja Noble}.}
  \bibinfo{year}{2018}\natexlab{}.
\newblock \bibinfo{booktitle}{\emph{{Algorithms of Oppression}}}.
\newblock \bibinfo{publisher}{New York University Press}.
\newblock


\bibitem[\protect\citeauthoryear{O'Neill}{O'Neill}{2017}]%
        {oneill}
\bibfield{author}{\bibinfo{person}{Cathy O'Neill}.}
  \bibinfo{year}{2017}\natexlab{}.
\newblock \bibinfo{title}{The Ivory Tower Can't Keep Ignoring Tech}.
\newblock
\newblock
\urldef\tempurl%
\url{https://www.nytimes.com/2017/11/14/opinion/academia-tech-algorithms.html}
\showURL{%
Retrieved January 3, 2019 from \tempurl}


\bibitem[\protect\citeauthoryear{Pichai}{Pichai}{2018}]%
        {GooglePrin}
\bibfield{author}{\bibinfo{person}{Sundar Pichai}.}
  \bibinfo{year}{2018}\natexlab{}.
\newblock \bibinfo{title}{AI at Google: our principles}.
\newblock
\newblock
\urldef\tempurl%
\url{https://www.blog.google/technology/ai/ai-principles/}
\showURL{%
Retrieved December 19, 2018 from \tempurl}


\bibitem[\protect\citeauthoryear{Rader, Cotter, and Cho}{Rader
  et~al\mbox{.}}{2018}]%
        {Rader2018}
\bibfield{author}{\bibinfo{person}{Emilee Rader}, \bibinfo{person}{Kelley
  Cotter}, {and} \bibinfo{person}{Janghee Cho}.}
  \bibinfo{year}{2018}\natexlab{}.
\newblock \showarticletitle{{Explanations as Mechanisms for Supporting
  Algorithmic Transparency}}. In \bibinfo{booktitle}{\emph{the 2018 CHI
  Conference}}. \bibinfo{publisher}{ACM Press}, \bibinfo{address}{New York, New
  York, USA}, \bibinfo{pages}{1--13}.
\newblock


\bibitem[\protect\citeauthoryear{Robertson, Lazer, and Wilson}{Robertson
  et~al\mbox{.}}{2018}]%
        {Robertson2018}
\bibfield{author}{\bibinfo{person}{Ronald~E Robertson}, \bibinfo{person}{David
  Lazer}, {and} \bibinfo{person}{Christo Wilson}.}
  \bibinfo{year}{2018}\natexlab{}.
\newblock \showarticletitle{{Auditing the Personalization and Composition of
  Politically-Related Search Engine Results Pages}}. In
  \bibinfo{booktitle}{\emph{the 2018 World Wide Web Conference}}.
  \bibinfo{publisher}{ACM Press}, \bibinfo{address}{New York, New York, USA},
  \bibinfo{pages}{955--965}.
\newblock


\bibitem[\protect\citeauthoryear{Ross and Swetlitz}{Ross and Swetlitz}{2018}]%
        {Ross2018}
\bibfield{author}{\bibinfo{person}{Casey Ross} {and} \bibinfo{person}{Ike
  Swetlitz}.} \bibinfo{year}{2018}\natexlab{}.
\newblock \bibinfo{title}{IBM’s Watson Recommended ‘unsafe and Incorrect’
  Cancer Treatments}.
\newblock
\newblock
\urldef\tempurl%
\url{https://www.statnews.com/2018/07/25/ibm-watson-recommended-unsafe-incorrect-treatments/}
\showURL{%
Retrieved December 17, 2018 from \tempurl}


\bibitem[\protect\citeauthoryear{Singh and Ghosh}{Singh and Ghosh}{2017}]%
        {Singh2017}
\bibfield{author}{\bibinfo{person}{Vivek~K. Singh} {and} \bibinfo{person}{Isha
  Ghosh}.} \bibinfo{year}{2017}\natexlab{}.
\newblock \showarticletitle{Inferring Individual Social Capital Automatically
  via Phone Logs}.
\newblock \bibinfo{journal}{\emph{Proc. ACM Hum.-Comput. Interact.}}
  \bibinfo{volume}{1}, \bibinfo{number}{CSCW}, Article \bibinfo{articleno}{95}
  (\bibinfo{date}{Dec.} \bibinfo{year}{2017}), \bibinfo{numpages}{12}~pages.
\newblock
\showISSN{2573-0142}
\urldef\tempurl%
\url{https://doi.org/10.1145/3134730}
\showDOI{\tempurl}


\bibitem[\protect\citeauthoryear{Technologies}{Technologies}{2018}]%
        {UnityPrin}
\bibfield{author}{\bibinfo{person}{Unity Technologies}.}
  \bibinfo{year}{2018}\natexlab{}.
\newblock \bibinfo{title}{Introducing Unity's Guiding Principles for Ethical
  AI}.
\newblock
\newblock
\urldef\tempurl%
\url{https://blogs.unity3d.com/2018/11/28/introducing-unitys-guiding-principles-for-ethical-ai/}
\showURL{%
Retrieved December 19, 2018 from \tempurl}


\bibitem[\protect\citeauthoryear{Wakabayashi}{Wakabayashi}{2018}]%
        {Wakabayashi2018}
\bibfield{author}{\bibinfo{person}{Daisuke Wakabayashi}.}
  \bibinfo{year}{2018}\natexlab{}.
\newblock \bibinfo{title}{Self-Driving Uber Car Kills Pedestrian in Arizona,
  Where Robots Roam}.
\newblock
\newblock
\urldef\tempurl%
\url{https://www.nytimes.com/2018/03/19/technology/uber-driverless-fatality.html}
\showURL{%
Retrieved December 17, 2018 from \tempurl}


\bibitem[\protect\citeauthoryear{Wexler}{Wexler}{2018}]%
        {wexler2018life}
\bibfield{author}{\bibinfo{person}{Rebecca Wexler}.}
  \bibinfo{year}{2018}\natexlab{}.
\newblock \showarticletitle{Life, liberty, and trade secrets: Intellectual
  property in the criminal justice system}.
\newblock \bibinfo{journal}{\emph{Stan. L. Rev.}}  \bibinfo{volume}{70}
  (\bibinfo{year}{2018}), \bibinfo{pages}{1343}.
\newblock


\end{thebibliography}
